\documentclass[draf]{aa}
\usepackage{natbib}
\usepackage{color}
\usepackage{graphicx}
\usepackage[varg]{txfonts}
\usepackage{hyperref}
\hypersetup{citecolor=black, urlcolor=blue, linkcolor=black, colorlinks=true}

\date{Version: \today}

\makeatletter
\renewcommand{\fnum@figure}{{Figure \thefigure}}
\renewcommand{\fnum@table}{{Table \thetable}}
\makeatother

\newcommand{\fig}[1]{Fig.\,\ref{#1}}
\newcommand{\figs}[1]{Figs.\,\ref{#1}}

\newcommand{\eqn}[1]{Equation\,(\ref{#1})}

\title{Magnetic reconnection sustains the mass budget of the solar wind}

\titlerunning{mass budegt}

\author{Yajie Chen \inst{1}
          \and
          Hardi Peter \inst{1,2}
          \and
          Damien Przybylski  \inst{1}
          \and
          Haruhisa Iijima \inst{3}
          \and
          Lakshmi Pradeep Chitta  \inst{1}
          }

\institute{ %1%
            Max-Planck Institute for Solar System Research (MPS),
            37077 G\"{o}ttingen, Germany
            \email{cheny@mps.mpg.de}
            \and
            Institute for Solar Physics (KIS), 
            Georges-K{\"o}hler-Allee 401A, 
            79110 Freiburg, Germany
            \and
            Institute for Space-Earth Environmental Research, Nagoya University, Furocho, Chikusa-ku, Nagoya, 464-8601, Aichi, Japan
        }

\begin{document}

\abstract{
%\todo{[HI: Figure 7 may be impressive because the solar wind mass does not come from the solar surface, but come from the closed loops. The downward mass flux at the chromosphere is not expected at least in 1-D Alfvenic turbulence models from the photosphere (e.g., CvE07). Thus, it may be worth emphasized in the abstract, as one of the main results.]}\newline
The solar wind originates from regions of open magnetic fields on the Sun, but the relevant processes remain unsolved. We present a self-consistent numerical model of the source region of the wind, in which jets similar to those observed on the Sun naturally emerge due to magnetic reconnection between closed and open magnetic fields. In this process material is transferred from closed to open field lines and fed into the solar wind. We quantify the mass flux through the magnetic field connected to the heliosphere and find that it greatly exceeds the amount required to sustain the wind. This {supports} a decades-old suspicion based on spectroscopic observations and shows that magnetic reconnection in the low solar atmosphere {could} sustain the solar wind.
}

\maketitle

\section{Introduction}

The solar wind is a continuous flow of charged particles from the hot atmosphere of the Sun, the corona \citep{1958ApJ...128..664P,1962Sci...138.1095N}, and shapes the heliosphere and governs the space weather environment \citep{1961PhRvL...6...47D,2013LRSP...10....3P,2007LRSP....4....1P}. This wind originates in regions where the Sun’s magnetic field opens into interplanetary space \citep{2009LRSP....6....3C,2016SSRv..201...55A}.% yet how plasma from the solar atmosphere is loaded onto these open field lines remains unresolved. 
%Traditional models assume a continuous outflow from the surface into open flux tubes \citep{1958ApJ...128..664P,2007ApJS..171..520C}.
While modern multi-spacecraft observations give some access to the source region(s) of the solar wind, the actual physical mechanisms that generate and accelerate the wind remain uncertain \citep[][]{2010ApJ...711.1044R,2010ApJ...720..824C,2020JGRA..12526005V}. Two of the most accepted hypotheses are based on the dissipation of magnetohydrodynamic (MHD) waves on open magnetic fields \citep[e.g.,][]{1986JGR....91.4111H,2007ApJS..171..520C,2007Sci...318.1574D,2011Natur.475..477M,2013ApJ...778..176O,2019ApJ...880L...2S,2021ApJ...907...55M} and on magnetic reconnection between closed and open field lines, i.e., interchange reconnection \citep[e.g.,][]{1992sws..coll....1A,1999JGR...10419765F,2020ApJ...904..199W,2023ApJ...945...28R,2023Natur.618..252B}.

The potential key role of interchange reconnection was derived from spectroscopic investigations of the source region of the solar wind: the observation of downward flows of cooler plasma spatially below hotter plasma showing an upward flow into the corona. Based on this, it has been proposed that interchange reconnection drives upflows above heights of ca. 5 Mm and downflows below ca. 5 Mm above the solar surface \citep{2005Sci...308..519T,2013ApJ...770....6Y,2025ApJ...985...27U}. Recent self-consistent MHD models have started to investigate the roles of interchange reconnection \citep{2023ApJ...951L..47I} in supplying energy to the solar wind, however they did not evaluate how well these models reproduce events in the source region, nor did they quantify the contributions of interchange reconnection in the mass supply to open field lines, i.e. the solar wind.

In this study, we use a self-consistent 3D MHD simulation of the source region of the wind and {compare it with}
%validate it by comparing with
coronal observations. By this we
%can confirm and quantify
{could provide evidence for and estimate}
the role of interchange reconnection in the mass supply to the nascent solar wind.

\section{MHD simulations and observations}

%>>>>>>>>>>>>>>>>>>>>>>>>>>>>>>>>>>>>>>>>>>>>>>>>>>>>>>>>>>>>>>>>>>>>>>>>>>>>>>
\begin{figure*}[ht]
\sidecaption
\centering {\includegraphics[width=10.5cm]{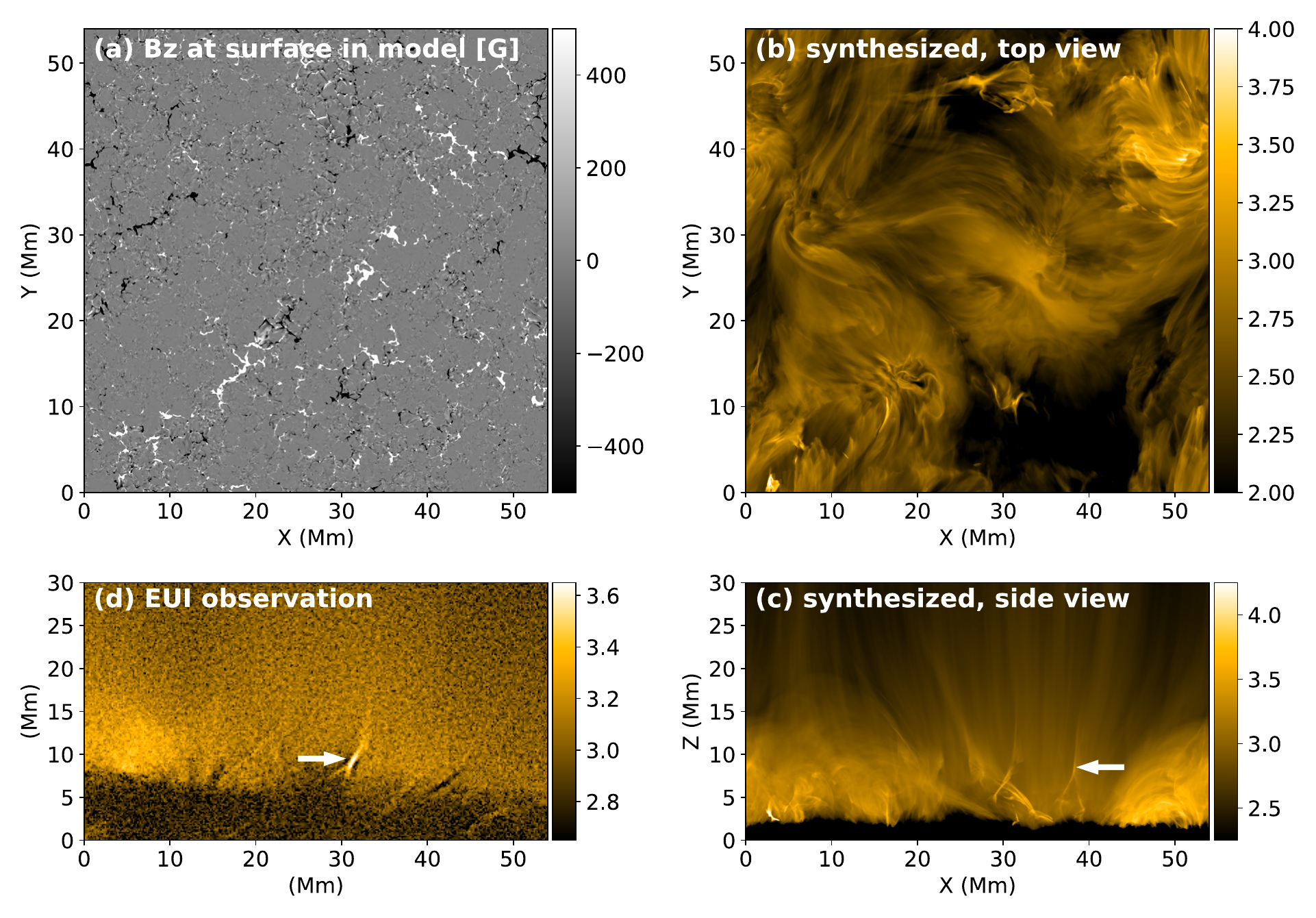}} 
\caption{{Modeled source region of the solar wind and comparison to observations.}
(a) Vertical component of magnetic field in the photosphere of the 3D MHD  model.
%(white and black show opposite magnetic polarities, saturated at $\pm$500 G).
(b) Coronal emission synthesized from the model in the 174 {\AA} passband integrated along the vertical direction, corresponding to an observation near the center of the solar disk.
(c) Same as (b) but integrated along the horizontal (Y) direction, corresponding to an off-limb observations.
(d) Coronal observation of the source region of the wind at the limb with HRI${\rm{_{EUV}}}$ onboard Solar Orbiter at 174 {\AA} (at 04:49:21 UT on 2022 March 30).
The white arrows mark the location of (inverse-Y-type) jets, one example in the observation and a comparable jet in the model.
The field of view of model and observation in (c) and (d) is the same.
The intensity maps are shown on a logarithm scale with arbitrary units.
An animation is available online.
%\textcolor{red}{HP: add an arrow to point to a jet in the model in panel (c).}
%\textcolor{red}{I would suggest to exchange panels (c) and (d), still keeping the numbering scheme, so that a-d is ordered in a clockwise circular fashion. Then the simulated side view is below the observations. Also in panel d you should add "EUI observation" and in panel c "synthesized, side view" and in b: "synthesized, top view". In panel a you could add "vertical field at surface in model".}
}
\label{fig:overview}
\end{figure*}
%<<<<<<<<<<<<<<<<<<<<<<<<<<<<<<<<<<<<<<<<<<<<<<<<<<<<<<<<<<<<<<<<<<<<<<<<<<<<<<

The numerical model of the solar wind source covers a computational domain extending from the upper convection zone below the solar surface to the lower corona, solving the full set of radiative MHD equations (see Appendix \ref{sect:Simulations}).
The self-consistently generated photospheric field exhibits large-scale network patterns with kilo-Gauss field strengths along with small-scale salt-and-pepper internetwork fields (see \fig{fig:overview}a), consistent with observations \citep{2019LRSP...16....1B}. To account for the magnetically open nature of the magnetic field, a uniform vertical field with 5\,G strength has been added at the initial condition.
The simulation self-consistently maintains a 1~MK corona and generates a continuous outflow, sustaining a mass flux of 2.7$\times$10$^{-10}$~g~cm$^{-2}$~s$^{-1}$, matching the required values for mass loss rates to sustain the solar wind \citep{1977ARA&A..15..363W}. 
%\\ \todo{But what about the open field? You simply add 5 G? this needs a half sentence. I added a statement above, but I am not sure if this is correct.}

To compare our simulations with observations, we synthesized extreme ultraviolet (EUV) emission in 174 {\AA} passband {(see Appendix \ref{sect:Simulations})} of the Extreme Ultraviolet Imager (EUI) \citep{EUI} onboard Solar Orbiter \citep{SolO}. The emission in this passband is mainly from $\sim$1~MK plasma.
Images were synthesized for both disk center and off-limb views (see \fig{fig:overview}(b--c)).
In the synthesized images bright closed loops appear at the corners of the on-disk view and at both sides of the off-limb image, representing coronal bright points \citep{2019LRSP...16....2M}.
{The magnetic connectivity in the simulations is discussed in Appendix~\ref{sect:connectivity}}.

The High Resolution Imager of EUI  (HRI$_{\rm{EUV}}$) at 174 {\AA} provides coronal images with unprecedented spatial and temporal resolution, making it ideal for capturing dynamics of jets at fine-scales. 
We analyzed HRI$_{\rm{EUV}}$ images of a coronal hole
%--a region that appears darker when observing the Sun in X-ray and EUV wavelengths and is a major source of the solar wind--
near the south pole of the Sun taken 
%from 04:30 to 05:00 UT 
on 2022 March 30 {(see Appendix \ref{sect:obs})}.
A zoomed-in snapshot of the observation taken at 04:49 UT is shown in \fig{fig:overview}(d).
The image reveals a coronal bright point with sizes of $\sim$10 Mm on the left side, closely resembling closed loops in our simulations.
Both, the synthesized coronal images and the  HRI$_{\rm{EUV}}$ observations, reveal small-scale jets. These jets are best visible in off-limb views due to projection effects.

\section{Jets as imprints of interchange reconnection}

%>>>>>>>>>>>>>>>>>>>>>>>>>>>>>>>>>>>>>>>>>>>>>>>>>>>>>>>>>>>>>>>>>>>>>>>>>>>>>>
\begin{figure*}[ht]
\centering {\includegraphics[width=14cm]{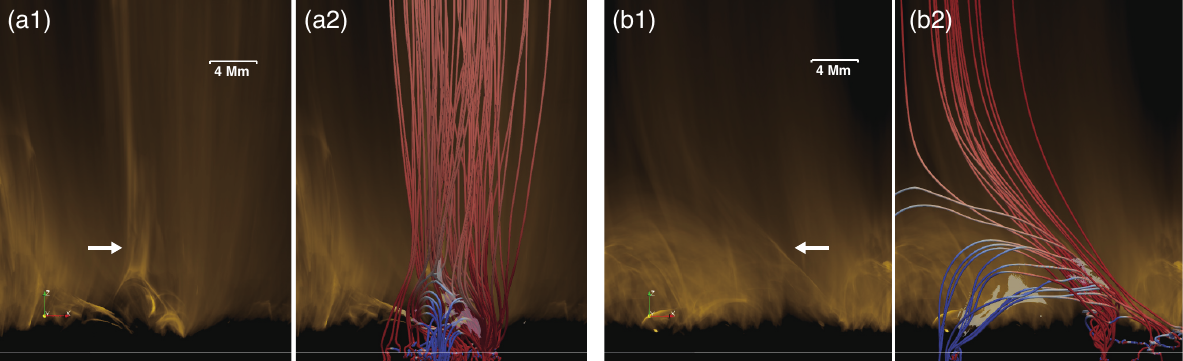}} 
\caption{
{Magnetic field structure in and around two coronal jets in the model.}
(a1) Synthesized 174 {\AA} image of a small field of view. The white arrow marks the location of a jet.
(a2) Same as (a1), but overlaid with magnetic field lines showing the magnetic field structure in and around the jet.
The field lines are colored by the vertical component of magnetic field with blue and red indicating negative and positive values.
The field line that are colored red only are (locally) open reaching the top boundary, while those changing in color from blue to red are field lines closing back to the surface.
(b1--b2) Same as (a1--a2) but for another jet.
The white straight lines indicates the location of the photosphere.
} %\textcolor{red}{HP: rename panels: a$\to$a1,  b$\to$a2, c$\to$b1 d$\to$b2}
\label{fig:magnetic_field_lines}
\end{figure*}
%<<<<<<<<<<<<<<<<<<<<<<<<<<<<<<<<<<<<<<<<<<<<<<<<<<<<<<<<<<<<<<<<<<<<<<<<<<<<<<

In {coronal} observations, jets with a shape of an inverted-Y are one of the most prominent structures in the coronal holes \citep{2016SSRv..201....1R}.
One such jet is identified at the center of the field of view (FOV) in \fig{fig:overview}(d). %\todo{I think you should use an arrow to mark the jet}
Similarly, the EUV images synthesized from the model reveal multiple Y-shaped jets.
They exhibit upward directed speeds of more than 100 km~s$^{-1}$, spatial scales along their length of a few megameters, widths of several hundred kilometers, and lifetimes of several hundred seconds. %\todo{mention size across/width?}
These properties are consistent with previous EUV and X-ray jets observed in coronal holes \citep{2016SSRv..201....1R,2023Sci...381..867C}.
The similarity between the simulated and observed jets suggests that the driving mechanism in the model is also representing the driver of the jets on the Sun.

To investigate the formation mechanism of the jets, we isolate a Y-shaped example in our simulations as shown in \fig{fig:magnetic_field_lines}(a1).
The jet has a lifetime of $\sim$5 minutes and recurs multiple times at the same location.
To investigate its magnetic structure and evolution, we traced magnetic field lines from 40 randomly selected points around it.
The magnetic field lines in and around the jet shown in \fig{fig:magnetic_field_lines}(a2) reveal that the jet originates from a reconnection site near its base, where nearly anti-parallel magnetic field lines reconnect--driven by flux emergence.
As the reconnection occurs, plasma is ejected along the newly formed open field lines, producing a fast and collimated jet.
This formation mechanism is consistent with classical models of jets driven by interchange reconnection \citep[e.g.,][]{1995Natur.375...42Y,2008ApJ...673L.211M,2016ApJ...822...18N}. 
This process is the reconnection of a closed and a open magnetic field line and naturally leads to the formation of the Y-shaped jets. 
In this process part of the formerly closed field line becomes part of an open field line. Hence, during this process mass can be transferred from closed to open magnetic field lines.

In addition to Y-shaped jets, some fainter jets in the observations exhibit a mostly linear morphology (\fig{fig:magnetic_field_lines}b1).
%They appear less structured than Y-shaped jets but still visible as collimated outflows.
Their speeds are comparable to Y-shaped jets but their intensity in EUV emission is lower.%, and it is more difficult to distinguish them from the background of coronal emission. 
Besides, the lifetimes of the linear jets (20 to 100 s) are shorter than those of the Y-shaped jets (300 to 600 s).%\citep{2023Sci...381..867C}.
We isolated a linear jet from our simulations as shown in \fig{fig:magnetic_field_lines}(b1) and examined its surrounding magnetic field structure.
Unlike the Y-shaped jet, which forms during reconnection between nearly anti-parallel fields, the linear jet originates from regions where open and closed field lines interact at small angles. After the reconnection, when the jet disappears in EUV emission, the field lines become more aligned.
This small-angle reconnection is also known as component reconnection \citep{1974JGR....79.1546S} and often expected to occur in open field configurations \citep{2016ApJ...825L...3T,2020JGRA..12526005V}.
As component reconnection generally releases less magnetic energy compared to reconnection involving anti-parallel field lines, {it produces less heating and ejects a smaller total mass, which makes} the linear jets fainter and more challenging to detect in observations.

%>>>>>>>>>>>>>>>>>>>>>>>>>>>>>>>>>>>>>>>>>>>>>>>>>>>>>>>>>>>>>>>>>>>>>>>>>>>>>>
\begin{figure}
\centering {\includegraphics[width=8cm]{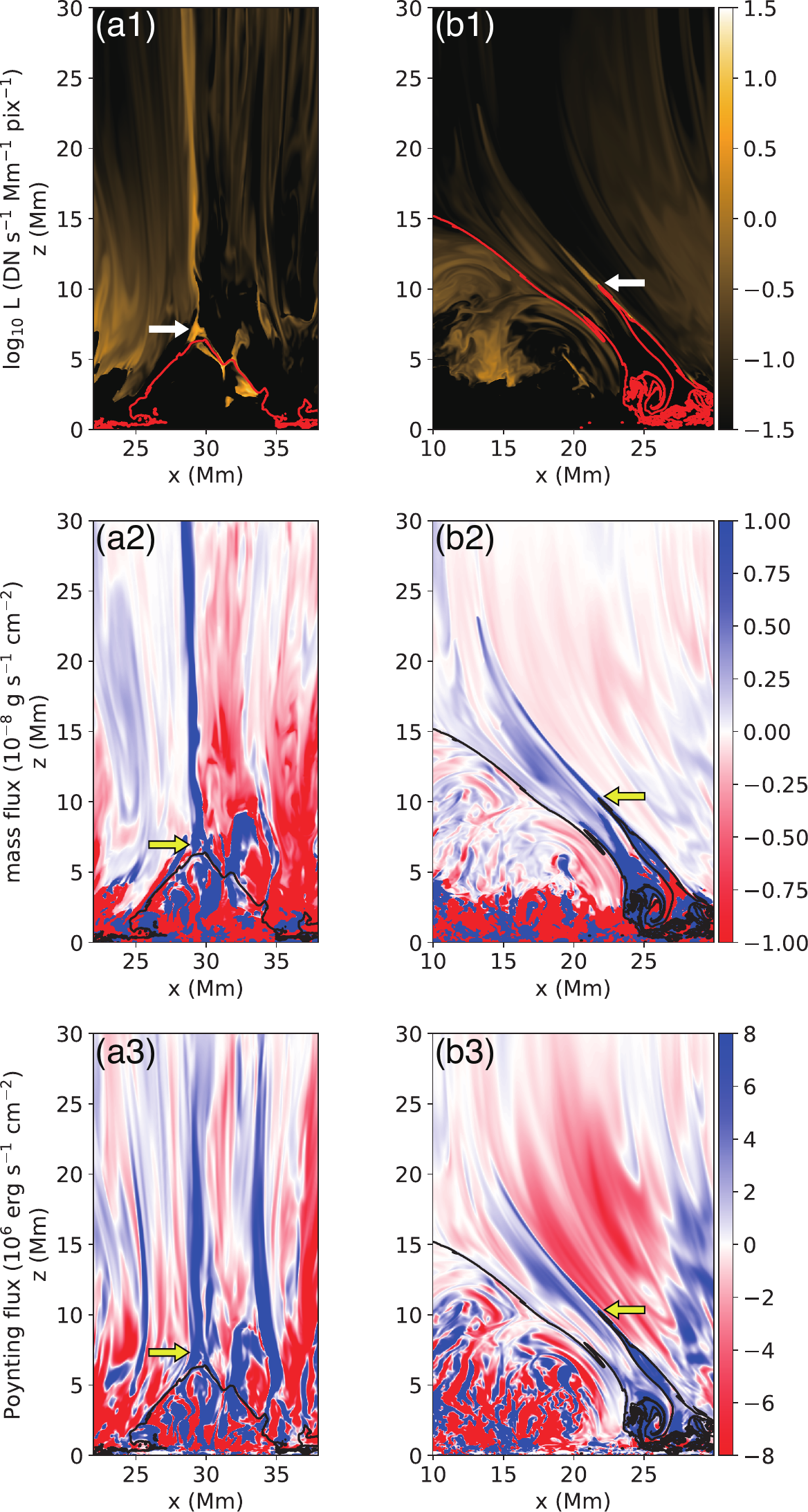}} 
\caption{
{Vertical slices through two jets.}
The left and right columns show the jets in panels (a) and (b) of \fig{fig:magnetic_field_lines}, respectively.
Panels (a1, b1) show the radiative losses per volume in the 174 {\AA} passband of EUI in a vertical slice. The red contours outline the boundaries between (locally) open and closed magnetic fields, with the open fields found above the respective contour.
The two jets form at the tip of the boundary between open and closed field (arrows).
Panels (a2, b2) show the mass flux and panels (a3, b3) the Poynting flux of magnetic energy. Blue indicate an upward directed flux. For both jets, there is an upward directed flux of mass and magnetic energy.
} 
%\textcolor{red}{%HP: Add arrows to the tip of the boundary of open and closed field, i.e. where the respective jet touches that boundary; 
%\newline
%Name panels: left column: (a1),(a2)(a3) and the right column (b1),(b2),(b3). \newline Then there is a direct correspondence to Fig 2.
%\newline
%Something is wrong with labels for the color tables (should be on far right, not on far left).}
\label{fig:vertical_slices}
\end{figure}
%<<<<<<<<<<<<<<<<<<<<<<<<<<<<<<<<<<<<<<<<<<<<<<<<<<<<<<<<<<<<<<<<<<<<<<<<<<<<<<

We confirm that both Y-shaped, and linear jets, originate near the boundaries of open and closed fields, typically at heights of 5--15 Mm (see \fig{fig:vertical_slices}).
The similarity in their magnetic topology implies that interchange reconnection (at either small or large angles) in the lower corona is the primary driver of these jets.
%As shown in \fig{fig:vertical_slices}, beyond serving as observational signatures of reconnection, 
%{\color{red} I find this new statement confusing. Since this is discussing the model, can you not just describe what you see? While you need to soften, or clarify, statements that relate to interpretation of the observations (i.e. in the discussion), here it does not seem necessary.
%In our model, the jets are signatures of reconnection. 
%The upwards propagating mass within them (as shown in \fig{fig:vertical_slices}) is from interchange reconnection, contributing to... }
{In our model, the jets are signatures of reconnection.
The upward propagating plasma within them (as shown in \fig{fig:vertical_slices}) arises from interchange reconnection, indicating that this process may contribute to the mass loading of open magnetic fields, and hence the origin of the solar wind.}

The speeds of the jets in our simulation reach only $\sim$200~km~s$^{-1}$, which is much less than the escape velocity from the Sun ($>$600~km~s$^{-1}$).
However, we find a simultaneous increase in Poynting flux carried by some of the jets in our model (\fig{fig:vertical_slices}).
This indicates that magnetic energy released during magnetic reconnection is transported upward along the newly opened field lines, likely in the form of transverse motions.
On average, a net upward Poynting flux of 0.9$\times$10$^5$~erg~cm$^{-2}$~s$^{-1}$ is sustained at 27 Mm (see \fig{fig:mass_energy_flux_cross}), 
which is comparable to the required energy loss estimated from in-situ observations \citep{2012SoPh..279..197L,2021A&A...650A..14L} assuming radial expansion. 
%%\todo{[HI:This value is 10x smaller than the requirement of Withbroe \& Noyes (1977). As LIM12 and LIM+21 does not account the super-radial expansion of magnetic flux tube. WN77 assumes the value of 6 (p.375).
%Sokolov+2013 suggests the required value of 1.1 x 10$^6$ W m$^{-2}$ T$^{-1}$ = 1.1 x 10$^5$ erg s$^{-1}$ cm$^{-2}$ G$^{-1}$. In this simulation, energy flux per flux imbalance is 0.18 x 10$^5$ erg s$^{-1}$ cm$^{-2}$ G$^{-1}$, which is 5x smaller than the requirement.
%It is better to evaluate the Poynting flux at the base of corona. According to Fig. 6, the value at z = 5 Mm is greater than 0.9 x 10$^5$ cgs. How about using the Poynting flux at z = 5 Mm for consistency with those rquirements?]} 
In addition, energy flux through Alfv\'enic waves in our model, estimated from the average root mean square horizontal velocity ($\sim$60 km s$^{-1}$) and electron density ($\sim$3.3$\times$10$^8$ cm$^{-3}$), is consistent with those estimated from spectroscopic observations in polar coronal holes \citep{2009A&A...501L..15B}. This energy flux would easily account to the energy requirements to power the solar wind accounting also for super-radial expansion in coronal holes.
%In addition, in our model we see an upward directed energy flux along the open field lines through outward propagating Alfv\'enic waves originating in the photosphere and chromosphere, with magnitudes comparable to those in earlier 3D MHD models \citep{2022A&A...665A.118F}. This additional flux of energy would easily account to the requirement to power the solar wind accounting also for super-radial expansion in coronal holes. This is consistent with an earlier scenario, where the wave power to accelerate the wind originates at the base of the open field lines \citep{2007ApJS..171..520C,2020ApJ...904..199W}.
%
%\todo{[HI: Instead of formulation in these literatures, YM Wang (2020 https://doi.org/10.3847/1538-4357/abbda6 ) discusses the energy flux requirement accounting the super-radial expansion.]}.
This wave energy will ultimately drive the acceleration of plasma, supporting the upward transport of mass flux higher into the corona and ultimately to the solar wind \citep{2024Sci...385..962R}. %The actual process of mass supply, however, can be quite different \citep{2020ApJ...904..199W}.

% 1.5 erg/cm2/s at 1AU ( Schwenn, 1991)         =  1.5 mW/m2
% radial expansion: 68 W/m2
% factor 7 super radial: 500 W/m2    (=5e5 erg/cm2/s)
% [consistent with Sololov et al 2013 for the avg 5 G in our model].
%
% Fig S3: 1e5 erg/cm2/s = 100 W/m2  net energy flux out of the top of the box
% not in Fig S2: 2000 W/m2 upward propagating wave flux reflected at top boundary.
% At bottom boundary: large wave flux going in (upwards) and out (downwards) without significant net effect.

\section{Mass supply through interchange reconnection}
\label{mass_supply_main}

%>>>>>>>>>>>>>>>>>>>>>>>>>>>>>>>>>>>>>>>>>>>>>>>>>>>>>>>>>>>>>>>>>>>>>>>>>>>>>>
\begin{figure}
\centering {\includegraphics[width=7.5cm]{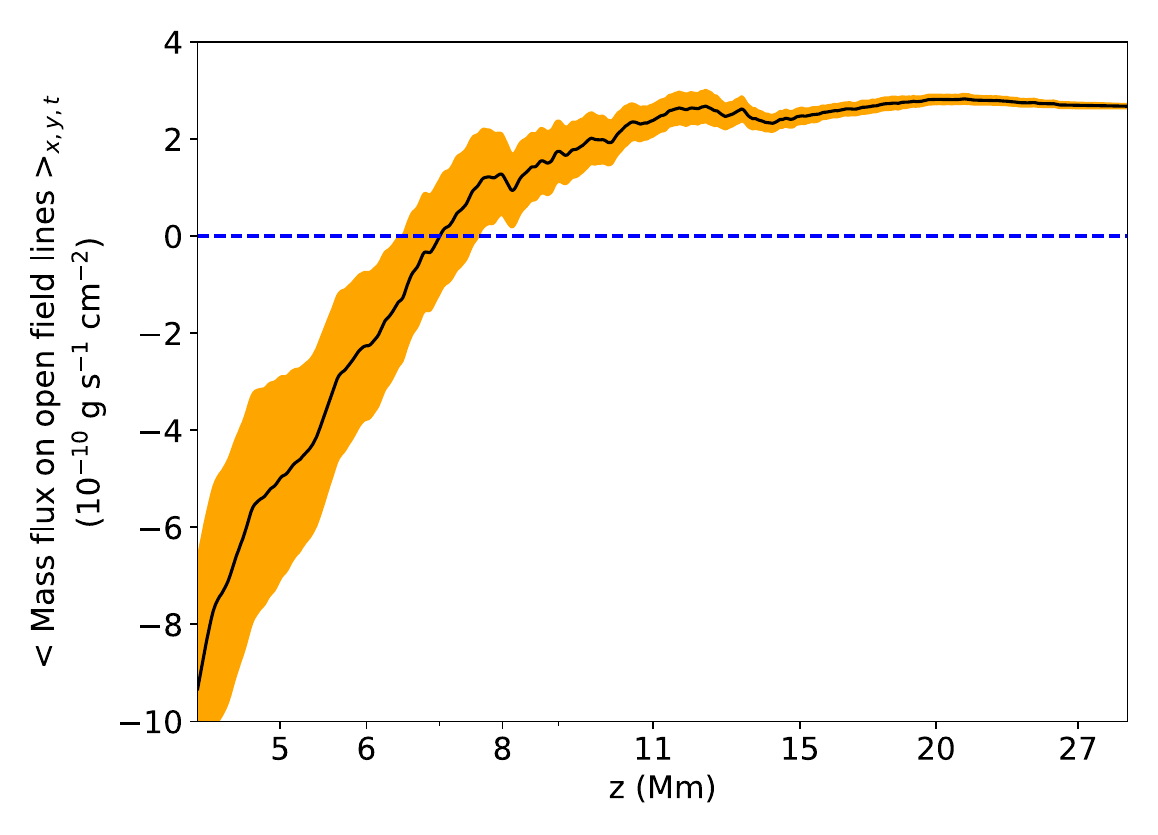}} 
\caption{
Mass flux in the magnetically open regions.
The black curve represents the mass flux as a function of height averaged over an $\sim$8-hour period and across the horizontal directions.
The yellow region indicates the temporal variability of the mass flux, {estimated as the standard deviations of different snapshots divided by the square root of the number of snapshots.}
The blue dashed line shows the zero flux.
%See \sect{mass_supply_main}.
On average, above $\sim$7~Mm the mass flows upwards, below it flows downwards. 
%The feeding of mass onto the open field lines at heights around $\sim$7~Mm is due to interchange reconnection with closed magnetic fields.
%\textcolor{red}{HP: Added these last two sentences starting "On average...". Strictly speaking, this is an interpretation of the figure that would not belong here but in the main text. However, because some people might look only at the figures, those people then would get the main message}
} 
\label{fig:mass_flux}
\end{figure}
%<<<<<<<<<<<<<<<<<<<<<<<<<<<<<<<<<<<<<<<<<<<<<<<<<<<<<<<<<<<<<<<<<<<<<<<<<<<<<<

The jets provide direct exemplarity evidence of interchange reconnection in coronal holes. However, also below the threshold of individually identifiable jets, continuous reconnection at the boundaries between open and closed magnetic field lines can inject mass into open magnetic fields and thus contribute to the nascent solar wind \citep{2012ApJ...758L..14R}.
Therefore, in the following we quantify the overall of mass flux transport from closed to open fields, rather than focusing on individual jet events.

To this end we first calculate the mass flux on the open magnetic fields averaged horizontally and over time at different heights.
As shown in \fig{fig:mass_flux}, we find a net downward mass flux below 5 Mm, indicating that plasma predominantly returns to the lower atmosphere at these heights.
However, above 20 Mm, where open field lines dominate our computational box, the net mass flux is upward, demonstrating that plasma continues to flow outward along open magnetic field lines and may escape as part of the solar wind.
Because of mass conservation, the mass transport from closed to open magnetic fields near the coronal base should play a key role in mass loading in the nascent solar wind.
These findings from our model compare well to spectroscopic observations \citep{2005Sci...308..519T}. However, in our model we can now investigate and quantify the nature of the mass supply.

To quantify the contribution of interchange reconnection in this process, we calculate the mass flux transported from the closed to open fields above 5 Mm (see Appendix \ref{sect:transport}).
We find that interchange reconnection injects a mass flux of 7.3$\times$10$^{-10}$\,g\,cm$^{-2}$\,s$^{-1}$ on average into open regions.
This is about 3.6 times higher than the well established value required to sustain the solar wind mass from coronal holes \citep{1977ARA&A..15..363W}.%, even when considering super-radial expansion in coronal holes. 
This suggests that interchange reconnection indeed plays a dominant role in mass loading of open magnetic fields, now finally quantifying the speculations based on spectroscopic observations \citep{2005Sci...308..519T}.

Interchange reconnection generates bidirectional flows on open field lines, with upflows driven by reconnection outflows and downflows caused by a combination of reconnection-driven downflows and plasma draining downward due to gravity. While downflows dominate at lower altitudes, the upflows persist at greater heights, feeding mass into the nascent solar wind.

\section{Summary and conclusions}

%We built a self-consistent 3D radiation MHD model of the source region of the solar wind and synthesized coronal images for a comparison with high-resolution EUV-imaging observations.
%Our simulation successfully reproduces small-scale coronal jets with Y-shaped and linear morphologies, exhibiting sizes, speeds, and lifetimes consistent with observations.
%These jets emerge at the boundaries between open and closed magnetic fields and are triggered by interchange reconnection.
%We find that interchange reconnection transports an average mass flux a few times higher than the required minimum flux to sustain the solar wind mass loss.
%This process naturally generates a net downward mass flux at and below the coronal base and net upward mass flux at greater heights, where open magnetic field lines dominate.
%Our results demonstrate that interchange reconnection is not only responsible for generating small-scale jets but also serves as a major mass supply mechanism for the nascent solar wind. 

%\todo{----------- HP: new summary incl. Parker wind suggested to replace above paragraph: --------}

The supply of material to the solar wind has been puzzling since the seminal description of the solar wind \citep{1958ApJ...128..664P}. In the original models the heliosphere is continuously connected to the solar surface through the magnetic field. 
Assuming mass conservation, the outflow speed in the photosphere (at optical depth unity) required to supply the mass flux to the solar wind would be $\leq$ 1~mm\,s$^{-1}$. Taking this speed at face value, the wind would need decades to cross the photosphere. 
Obviously, the high dynamics in the photosphere and chromosphere require that both the spatial complexity and temporal evolution be considered to understand how the Sun actually supplies the mass to the solar wind outflow. A natural place to look for the supply of mass is above the chromosphere, where the magnetic field is dominating energetics and hence can provide a more stable environment to host outflow channels.

Through our self-consistent 3D radiation MHD model we provide the explanation how this supply of mass is facilitated at the base of the corona by transferring material from magnetically closed regions into rapidly expanding coronal funnels that are magnetically open in the sense that they are connected to the heliosphere. In our model this is achieved through interchange reconnection that comes along with small-scale coronal jets, similar in morphology, size, speed, and lifetime to those abundantly seen in observations.
Hence our model is not only producing the right order of mass flux supplied to the solar wind, but it also matches the observed dynamics of the solar wind source region. This provides a comprehensive explanation of the supply of mass to the solar wind in its source region.
{Although our model does not include some effects such as non-equilibrium ionization or ion-neutral interactions, which may affect details, the overall picture should not change.}

% Simple radial expansion to tau_500=1 gives ca. 2e-6 m/s, i.e 2 microns/s.
% Even assuming a larger non-radial expansion of the wind of a factor of 10
% and a strong concentration of the magnetic flux tubes into the photosphere 
% with an area factor of 100 would give only 2 mm/s.
% At a speed of 1 mm/s the crossing time over 500 km
% (up to the temperature minimum) would be 15 years.

% Another way to estimate the velocity:
% nv/B is constant, where n is the number density and v is the velocity
% take the numbers in Fig 1(f) in \citep{2010ApJ...715L.121W} (measured at 1AU), nv/B is about 10^13 cm^{-2} s^{-1} G^{-1}
% in the photosphere, n~10^17 cm^{-3}. 
% assuming open field lines originate from magnetic network with B~10^3 G
% then v~0.1 cm/s
%
% HP: yes, this is an even faster way to do this. The non-radial expansion and the rapidly expanding funnels I mentioned above give an additional expansion by a factor of 1000 (in addition to the radial expansion). This is of course, directly related to the expansion of the magnetic field. Thanks for checking!

\begin{acknowledgements}
YC acknowledges funding provided by the Alexander von Humboldt Foundation.
The work of YC and DP was funded by the Federal Ministry for Economic Affairs and Climate Action (BMWK) through the German Space Agency at DLR based on a decision of the German Bundestag (Funding code: 50OU2201).
LPC gratefully acknowledges funding by the European Union (ERC, ORIGIN, 101039844).
Solar Orbiter is a space mission of international collaboration between ESA and NASA, operated by ESA. The EUI instrument was built by CSL, IAS, MPS, MSSL/UCL, PMOD/WRC, ROB, LCF/IO with funding from the Belgian Federal Science Policy Office (BELSPO/PRODEX PEA 4000112292 and 4000134088); the Centre National d’Etudes Spatiales (CNES); the UK Space Agency (UKSA); the Bundesministerium für Wirtschaft und Energie (BMWi) through the Deutsches Zentrum für Luft- und Raumfahrt (DLR); and the Swiss Space Office (SSO). 
We gratefully acknowledge the computational resources provided by the Cobra and Raven supercomputer systems of the Max Planck Computing and Data Facility (MPCDF) in Garching, Germany.
\end{acknowledgements}

\bibliography{sample}{}
\bibliographystyle{aa}

\begin{appendix}
\onecolumn

\section{Simulations} 
\label{sect:Simulations}
We used the coronal extension of the MURaM code \citep{MURaM,MURaM2017} to construct our model.
The computational domain spans 54 Mm $\times$ 54 Mm horizontally and extends from $\sim$20 Mm below to $\sim$30 Mm above the solar surface.
The grid size is $\sim$52.7 km in the horizontal directions and 20 km in the vertical direction.
Periodic boundary conditions are applied in the horizontal directions.
The bottom boundary allows both inflows and outflows with symmetric magnetic field components.
The upper boundary permits only outflows, and the magnetic field is prescribed by a potential boundary condition.

The simulation was initialized from a snapshot of a quiet Sun simulation without net magnetic flux imbalance, where the upper boundary was $\sim$500 km above the surface.
{In the model, a small-scale dynamo \citep{2023SSRv..219...36R} self-consistently generates magnetic field in the convection zone and photosphere.}
A 5 G vertical magnetic field was added at each grid cell, and the simulation was run for 6.1 hours.
The model was then extended to 6 Mm above the surface and evolved for additional 4 hours to establish a transition region.
Subsequently, we expanded the domain by another 24 Mm in the vertical direction and ran it for $\sim$5.8 hours until it stabilized.
Afterward, we continued the simulation for another $\sim$8.3 hours where we write out the state of the computational domain with a cadence of $\sim$5.5 minutes.
Additionally, to study the evolution of magnetic field structures around the two jets in \fig{fig:magnetic_field_lines}, we restarted the simulation from two snapshots with the same setup to generate two datasets with a higher writeout cadence of 10~seconds over 20~minutes.

Although our simulation does not extend into the solar wind regions, the domain size is sufficient to investigate the role of interchange reconnection in mass transport in the nascent solar wind. 
Moreover, the high spatial resolution allows us to resolve granulation in the photosphere and capture small-scale coronal dynamics, including jets with sizes comparable to those observed by HRI${_{\rm{EUV}}}$ on Solar Orbiter \citep[e.g.,][]{2021A&A...656L..13C,2021ApJ...918L..20H,2022A&A...664A..28M,2023Sci...381..867C}. %\todo{[HP: I would add here some citations, e.g. from Sudip and Pradeep. Yes, they have been mentioned in the main text, but it does not hurt to cite them here, too.]}
Essentially, our setup is similar to the model used in \citep{2023ApJ...951L..47I} from the convection zone to the lower corona, where an Alfv\'enic slow solar wind was produced.
Considering that our simulation satisfies the observational requirements for mass and energy flux at coronal heights, we are confident on producing a realistic solar wind parameters by extending our model to tens of solar radii.

To compare with observations, we synthesized coronal emission in the EUI 174 {\AA} passband, {dominated by the Fe~{\sc{ix}} (171.1 {\AA}) and Fe~{\sc{x}} (174.5, 177.2 {\AA}) lines}, following the method in \citep{2021A&A...656L...7C}. 
The response function $G(n_e,T)$, which depends on electron density and temperature and peaks around 1~MK, was used to calculate radiative loss as $n_e^2G(n_e,T)$ at each grid cell.
{$G(n_e,T)$ was obtained from CHIANTI version 10.0 \citep{Chianti,Chianti-v10} using coronal abundances %\citep{Feldman1992} 
and the standard CHIANTI ionization equilibrium. {The calculation was based on the spectral response \citep[updated from][]{EUI}} and is described in detail by \citet{2021A&A...656L...7C}.}
Since coronal emission is optically thin, we integrated the radiative losses along the vertical and horizontal directions to simulate disk center and off-limb observations, respectively.
{The absorption from hydrogen and helium {\citep[cf.][]{2022ApJ...938...60M}} was not considered in our study.}
{We kept the original spatial resolution of the model in synthesized image and did not take into account the spatial resolution of HRI$_{\rm{EUV}}$.}

\section{Magnetic topology and connectivity in the simulations}
\label{sect:connectivity}

%>>>>>>>>>>>>>>>>>>>>>>>>>>>>>>>>>>>>>>>>>>>>>>>>>>>>>>>>>>>>>>>>>>>>>>>>>>>>>>
\begin{figure*}[ht]
\centering {\includegraphics[width=\textwidth]{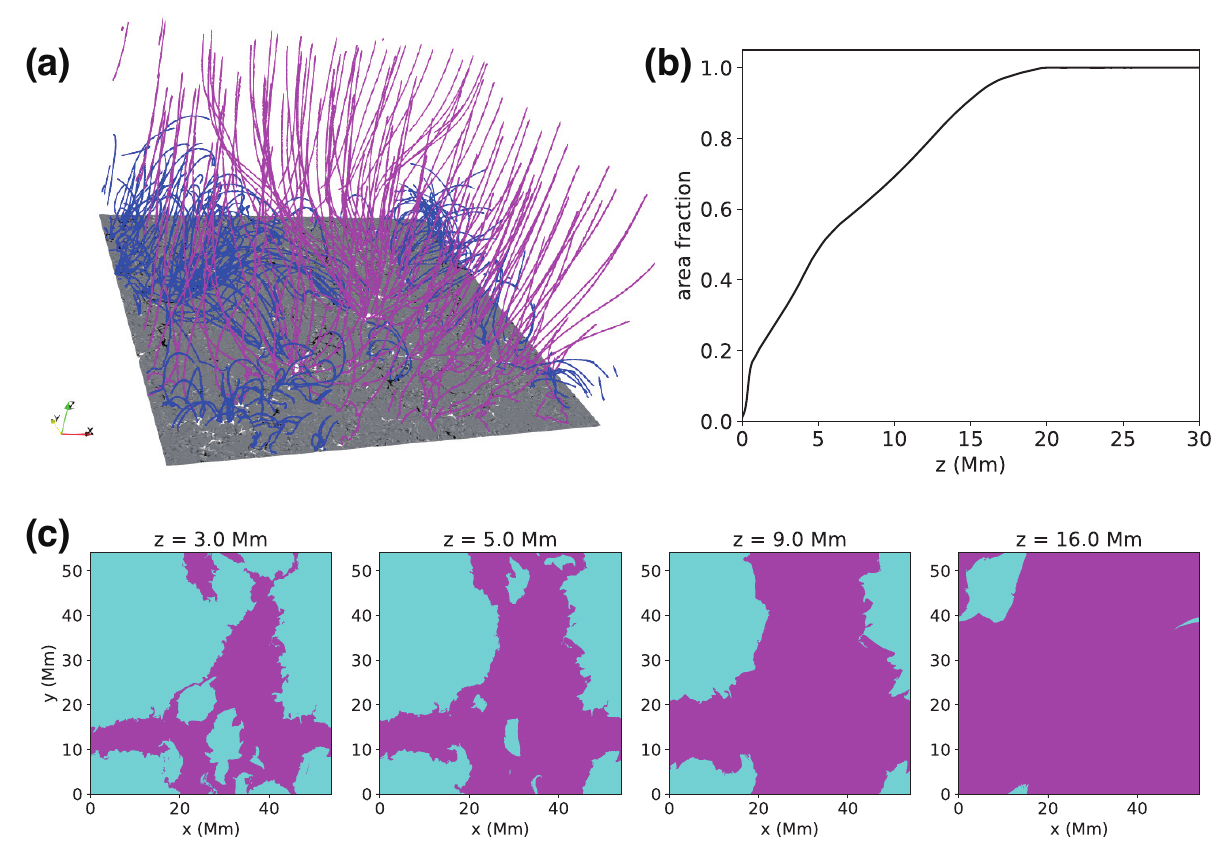}} 
\caption{
{Magnetic topology and connectivity in the model.}
In panel (a) the background image at the bottom shows the vertical component of the magnetic field in the photosphere (black and white indicate opposite polarities).
Blue lines represent closed field lines, while purple lines indicate open field lines, i.e. field lines that reach the top of the computational domain and might close back out in the outer solar atmosphere or in the heliosphere.
Panel (b) displays the fraction of area in a horizontal slice covered by open magnetic fields as a function of height for the snapshot shown in panel (a).
The panels (c) show the magnetic connectivity at heights of 3, 5, 9, and 16~Mm.
Blue and purple regions correspond to closed and open magnetic fields, respectively.
%\textcolor{red}{HP: changed Cyan to Blue in last sentence. I think Cyan is a light blue and this way it fits better to the color scheme for the field lines in panel (a).}
} 
\label{fig:open_region}
\end{figure*}
%<<<<<<<<<<<<<<<<<<<<<<<<<<<<<<<<<<<<<<<<<<<<<<<<<<<<<<<<<<<<<<<<<<<<<<<<<<<<<<

To investigate the large-scale magnetic field structures in our simulations, we present field lines traced from 625 uniformly distributed points at a height of 5 Mm in \fig{fig:open_region}(a).
The photospheric magnetogram is also shown in \fig{fig:open_region}(a).
In addition, \fig{fig:open_region} presents the area fraction of open regions as a function of height in the atmosphere and horizontal slices of the filling factor at different heights, illustrating the spatial distribution of open and closed magnetic fields, {to be compared with observations from extrapolations in the future.}

Our simulations reveal two types of closed magnetic loops:
large-scale loops extending up to 20 Mm, which appear at the corners of the computational domain and persist throughout the simulation, 
and small-scale newly emerged closed loops, which intermittently reach above 5 Mm.
These smaller loops are often associated with Y-shaped jets in the synthesized EUV images and have lifetimes typically shorter than one hour.

Furthermore, the boundaries between open and closed regions exhibit a complex and fractal structure, driven by turbulent convective motions in the photosphere \citep{2012ApJ...758L..14R}.
As a result, interchange reconnection continuously occurs at the open-closed field boundaries, driving persistent outflows.
The open magnetic field lines are rooted in narrow lanes in the photosphere and chromosphere, aligning with strong magnetic field concentrations.
Such a connection is consistent with the observations, which indicate that the nascent solar wind outflows originate from the magnetic network \citep{1999Sci...283..810H}.

\section{Observations}\label{sect:obs}

To relate our numerical model to the processes on the Sun, we use HRI$_{\rm{EUV}}$ observations of the south polar coronal hole taken on 2022 March 30 from 04:30 to 05:00 UT with a 3~s cadence.
The images have a plate scale of 0.492$^{\prime\prime}$ per pixel.
During these observations the instrument was 0.332 AU from the Sun, hence providing a spatial resolution of $\sim$237 km.
This study uses Level 2 data \citep{euidatarelease6}.
Since the jet activity in this dataset has been analyzed in detail already by \citep{2023Sci...381..867C}, we focus on a single image taken at 04:49 UT, which captures an evident Y-shaped off-limb jet.
%The jets' properties, including their width (a few hundred kilometers), speeds (over 100 km~s$^{-1}$), and lifetimes (tens to hundreds of seconds), are taken directly from their study.
To enhance the image contrast, we applied unsharp masking. For this, first the original image was smoothed over an 8$\times$8 pixel window. The smoothed version was then subtracted from the original image and the residual was amplified threefold before being added back, {which was applied in \fig{fig:overview}(d)}.

\section{Kinetic energies carried by the jets in the model}

We analyze the kinetic energies of the jets shown in \fig{fig:magnetic_field_lines}.
To simplify the calculation, we consider only the pixels with enhanced EUI 174 {\AA} emission and speeds exceeding 60~km~s$^{-1}$ at the time shown in \fig{fig:magnetic_field_lines}.
The velocity threshold is chosen as most small-scale coronal jets in our simulations reach speeds above this value.
This approach provides a lower limit on the kinetic energy carried by the jets, because some plasma is still accelerating, and part of it has already escaped from the computational domain.

We find that the kinetic energy of the linear jet is $\geq$3$\times$10$^{24}$\,erg, comparable to the typical energy of a nanoflare \citep{1988ApJ...330..474P}.
The Y-shaped jet carries a significantly higher kinetic energy of $\geq$2$\times$10$^{25}$\,erg, approximately an order of magnitude larger than that of the linear jet.
The energy difference is consistent with the stronger energy release expected from interchange reconnection in Y-shaped jets, where newly emerged closed loops reconnect with preexisting open fields, driving more energetic plasma ejections in general.

%The kinetic energy flux $E_k=\frac{1}{2}\rho v^2v_z$ at $z=8$ Mm is about 1.4$\times$10$^6$~erg~cm$^{-2}$~s$^{-1}$ within the linear jet and 1.7$\times$10$^6$~erg~cm$^{-2}$~s$^{-1}$ within the Y-shaped jet.

\section{Mass and Poynting Flux Transport in Open Magnetic Fields}
\label{sect:transport}

%\subsection*{Mass transport in open magnetic fields}

%>>>>>>>>>>>>>>>>>>>>>>>>>>>>>>>>>>>>>>>>>>>>>>>>>>>>>>>>>>>>>>>>>>>>>>>>>>>>>>
\begin{figure*}
\centering {\includegraphics[width=14cm]{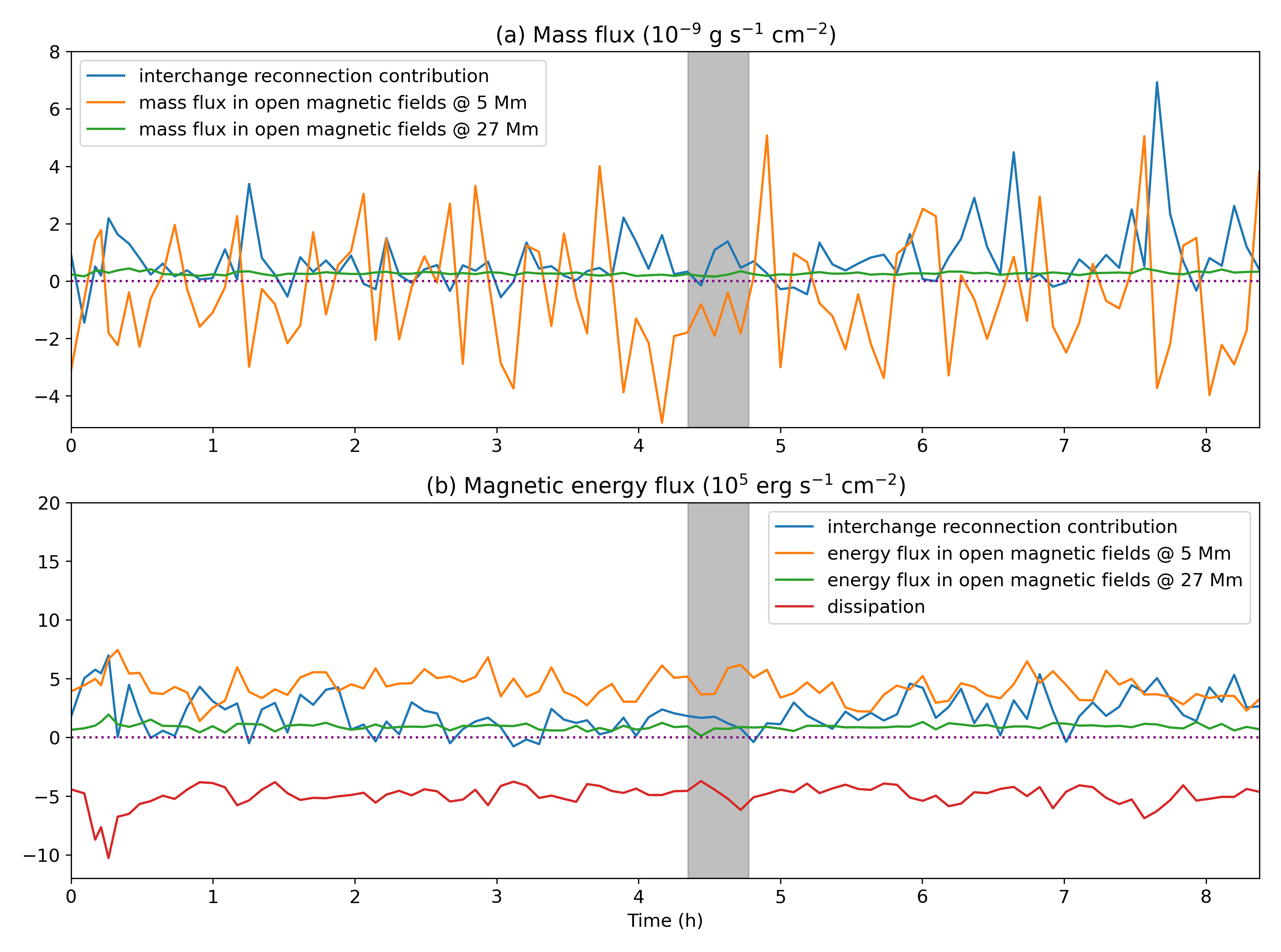}} 
\caption{
{Temporal evolution of the contributions to the mass and energy balances.}
Panel (a) shows the balance for the mass flux as detailed in \eqn{eq:cross_mass}. The three curves show the mass flux at the bottom at 5~Mm, ${\mathcal{M}_0}$ (orange), near the top at 27~Mm, ${\mathcal{M}_1}$ (green), and the mass flux through interchange reconnection from closed to open regions, ${\mathcal{M}_R}$ (blue).
Positive values indicate net upward propagating flux (${\mathcal{M}_0}$, ${\mathcal{M}_1}$) and flux transport from closed to open regions (${\mathcal{M}_R}$).
In a similar way panel (b) shows the contribution to the magnetic energy flux balance as detailed in \eqn{eq:cross_energy} with ${\mathcal{S}_0}$ (orange), ${\mathcal{S}_1}$ (green), and ${\mathcal{S}_R}$ (blue) indicating the fluxes at the top, bottom and through interchange reconnection, respectively.
Also work done by the Lorenz force and dissipation change the magnetic energy. Hence, we add the respective term ${\mathcal{S}_D}$ (red) from \eqn{eq:cross_energy} to this plot, with negative values indicating a loss of magnetic energy.
The purple dotted lines in both panels indicate zero flux.
{The grey-shaded region indicates the period shown in \fig{fig:mass_energy_flux_cross_high_cad}.}
%\textcolor{red}{HP: I would also add panel indices here, i.e. (a) and (b).}
}
\label{fig:mass_energy_flux_cross}
\end{figure*}
%<<<<<<<<<<<<<<<<<<<<<<<<<<<<<<<<<<<<<<<<<<<<<<<<<<<<<<<<<<<<<<<<<<<<<<<<<<<<<<

%>>>>>>>>>>>>>>>>>>>>>>>>>>>>>>>>>>>>>>>>>>>>>>>>>>>>>>>>>>>>>>>>>>>>>>>>>>>>>>
\begin{figure*}
\centering {\includegraphics[width=14cm]{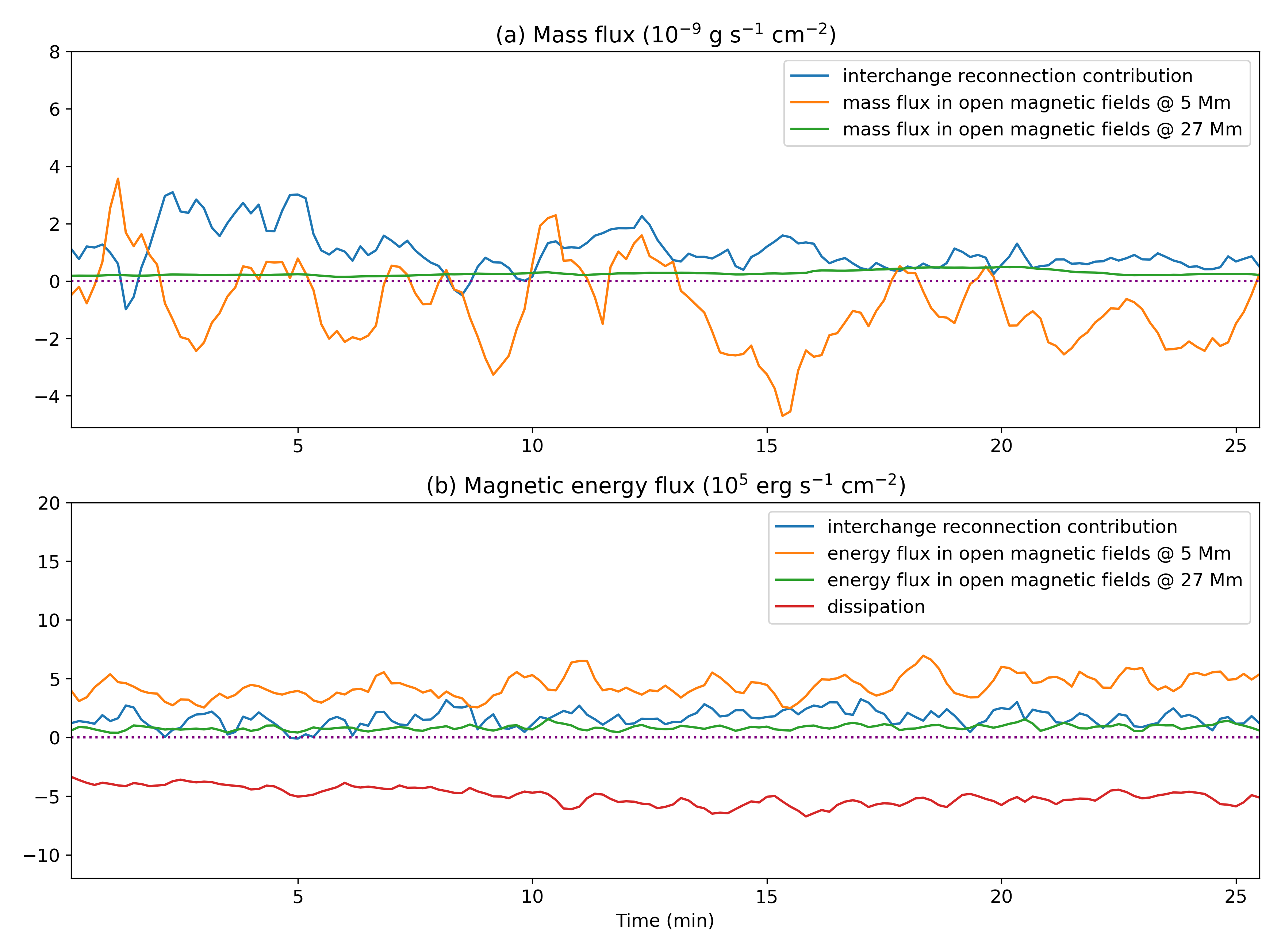}} 
\caption{
{Similar to \fig{fig:mass_energy_flux_cross} but for the high-cadence dataset, with the time range given by the grey-shaded region in \fig{fig:mass_energy_flux_cross}.}
}
\label{fig:mass_energy_flux_cross_high_cad}
\end{figure*}
%<<<<<<<<<<<<<<<<<<<<<<<<<<<<<<<<<<<<<<<<<<<<<<<<<<<<<<<<<<<<<<<<<<<<<<<<<<<<<<

%>>>>>>>>>>>>>>>>>>>>>>>>>>>>>>>>>>>>>>>>>>>>>>>>>>>>>>>>>>>>>>>>>>>>>>>>>>>>>>
\begin{figure*}
\centering {\includegraphics[width=\textwidth]{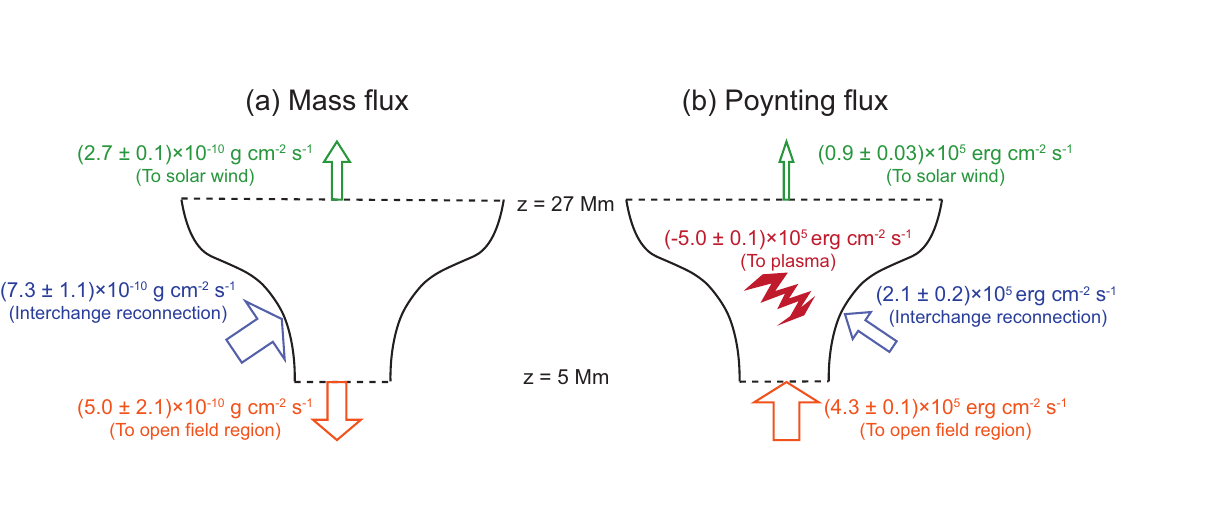}} 
\caption{{Schematic illustration of mass and energy flux transport in open magnetic fields.} The values of mass and energy flux quoted here are the long-term averages of the evolution shown in \fig{fig:mass_energy_flux_cross} for the different terms in \eqn{eq:cross_mass} and \eqn{eq:cross_energy}. The uncertainties quoted here represent the temporal variation across multiple snapshots of the model.
Panel (a) shows the mass flux at different heights ($\langle{\mathcal{M}_0}\rangle$ and $\langle{\mathcal{M}_1}\rangle$) and contributions from interchange reconnection ($\langle{\mathcal{M}_R}\rangle$). The {widths} of the arrows are indicative for the magnitude of the flux.
Panel (b) presents the Poynting flux at different heights ($\langle{\mathcal{S}_0}\rangle$ and $\langle{\mathcal{S}_1}\rangle$), with contributions from interchange reconnection ($\langle{\mathcal{S}_R}\rangle$) and changes due to Lorentz force work and Joule heating ($\langle{\mathcal{S}_D}\rangle$). 
The widths of the arrows are roughly scaled to illustrate relative magnitudes of the fluxes.
%\textcolor{red}{%\newline HP: Again, I would also add panel indices here, i.e. (a) and (b).
%\newline Is the size of the arrows indicative for the fluxes?}
}
\label{fig:cartoon}
\end{figure*}
%<<<<<<<<<<<<<<<<<<<<<<<<<<<<<<<<<<<<<<<<<<<<<<<<<<<<<<<<<<<<<<<<<<<<<<<<<<<<<<

%>>>>>>>>>>>>>>>>>>>>>>>>>>>>>>>>>>>>>>>>>>>>>>>>>>>>>>>>>>>>>>>>>>>>>>>>>>>>>>
\begin{figure*}
\centering {\includegraphics[width=14cm]{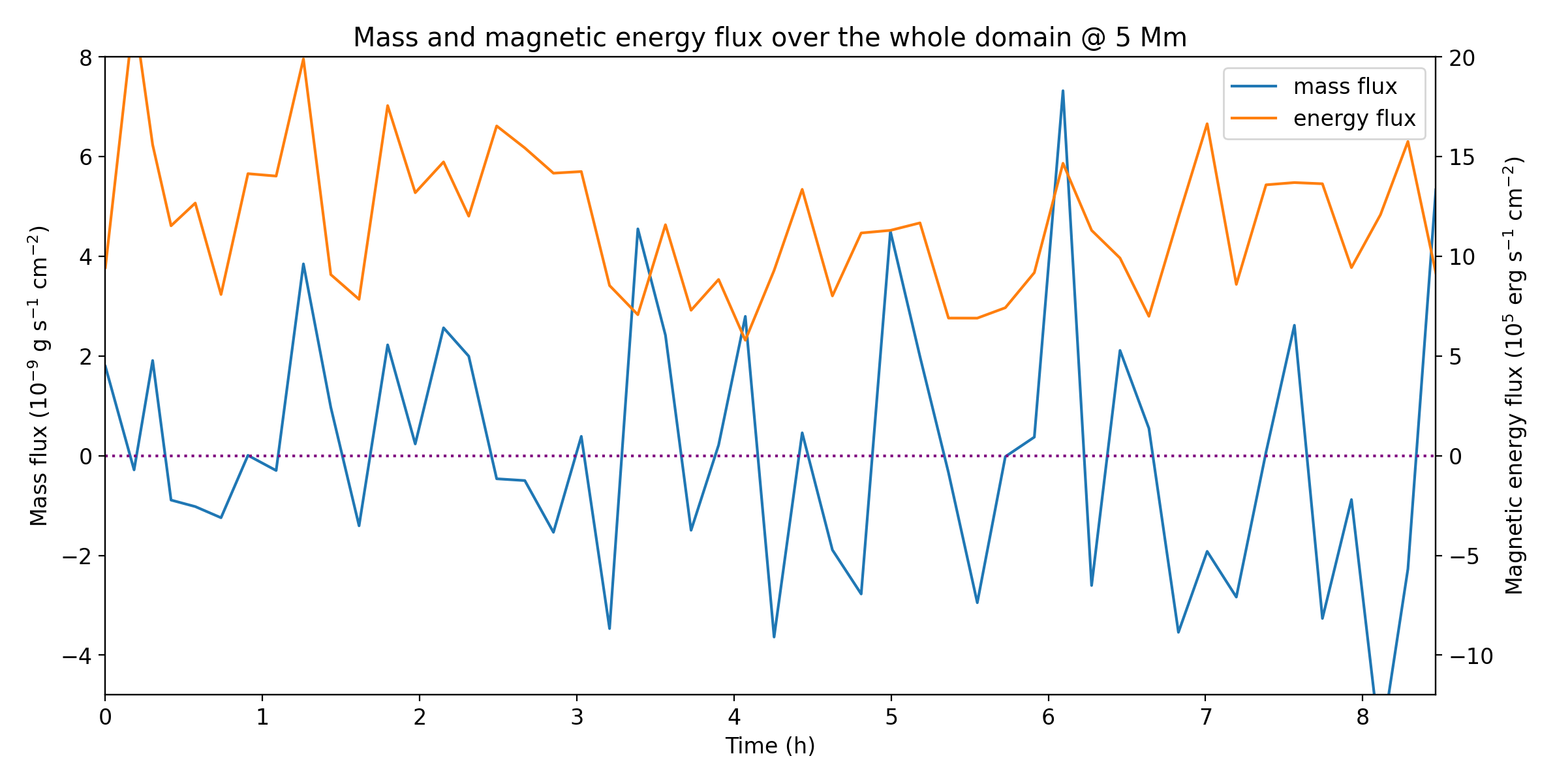}} 
\caption{
{Temporal evolution of mass (blue) and energy flux (orange) over the whole domain at 5 Mm for the low-cadence dataset.
Positive values correspond to net upward propagating flux.
The purple dotted lines indicate zero flux.}
}
\label{fig:flux_whole_domain_5Mm}
\end{figure*}
%<<<<<<<<<<<<<<<<<<<<<<<<<<<<<<<<<<<<<<<<<<<<<<<<<<<<<<<<<<<<<<<<<<<<<<<<<<<<<<

We used the Runge–Kutta–Fehlberg (RK45) method \citep{fehlberg1969klassische} to trace magnetic field lines in each snapshot.
To improve accuracy, we oversampled the horizontal directions by placing nine seed points per grid cell.
Field lines from these points were then integrated forward and backward.
To determine whether a grid cell was open or closed, we defined the filling factor $f$ as the fraction of field lines originating from the grid cell that reach the top boundary.
When we show the open-closed field boundaries in \figs{fig:vertical_slices} and \fig{fig:open_region}, grids where $f>0.5$ were labeled open, and those with $f\leq0.5$ were labeled closed.
The oversampling method reduces uncertainties in magnetic field tracing, especially at lower heights where the gradients of the magnetic field are large.

To calculate the mass transport in open magnetic fields, we rewrite the continuity equation to incorporate the filling factor:
\begin{equation*}
\nabla \cdot \left ( f\rho \mathbf{v} \right ) = \left ( \rho \frac{\partial}{\partial t}+\rho \mathbf{v} \cdot \nabla \right )f - \frac{\partial\left ( f\rho \right )}{\partial t}
\nonumber
\end{equation*}
where $\rho$ and $\mathbf{v}$ are density and velocity, respectively.
Integrating the equation over the horizontal domain from $z=z_0$ to $z=z_1$ gives:
\begin{equation}
\underbrace{\int_{A}\left ( f\rho v_z \right ) |_{z=z_1}dxdy}_{{\mathcal{M}_1}} - \underbrace{\int_{A}\left ( f\rho v_z \right ) |_{z=z_0}dxdy}_{{\mathcal{M}_0}} =  \underbrace{\int_{V}
 \left ( \rho \frac{\partial f}{\partial t} +\rho \mathbf{v} \cdot \nabla f \right ) dxdydz}_{\mathcal{M}_{R}} 
 - \displaystyle\int_{V} \frac{\partial\left ( f\rho \right )}{\partial t} dxdydz
\label{eq:cross_mass}
\end{equation}
%\todo{[HP: a suggestion: Maybe you could mark the terms in Eq (1)  in a fashion similar as I did now for the first term on the left-hand side. Then you can refer in the text much more easy and transparent to the different terms. This is then also useful for the discussion of Fig. 6 when you discuss the contribution of interchange connection and how it evolves in time. Same procedure for Eq. (2).]}\YC{YC: looks good to me. Now I have marked the terms in both equations, and I need to modify the text accordingly.}\\
where $v_z$ is the vertical component of velocity.
The terms ${\mathcal{M}_0}$ and ${\mathcal{M}_1}$ represent the net mass flux entering and leaving the open regions at heights of $z_0$ and $z_1$, respectively.
The term ${\mathcal{M}_R}$ accounts for the contribution of interchange reconnection from closed to open regions between $z_0$ and $z_1$.
The last term is the temporal changes of density in the open regions and is negligible after averaging over several hours.
The derivative is calculated with fourth order in space and second order in time.
To reduce the uncertainty from field line tracing, we chose $z_0=5$ Mm, as the filling factor is less disordered above this height compared to the chromosphere.
The choice of $z_1$ does not affect results as long as it is above 20 Mm, where all field lines are open and the mass flux remains nearly constant (\fig{fig:mass_flux}).
To reduce boundary effects, we chose $z_1 = 27$ Mm.

%\subsection*{Magnetic energy transport in open magnetic fields}

Interchange reconnection contributes to both mass and energy flux in the nascent solar wind.
While this study mainly focuses on mass transport, we also evaluate its contribution to magnetic energy flux following the procedure outlined in \citep{2023ApJ...951L..47I}.
The flux of magnetic energy is described by:
\begin{equation}
\underbrace{\int_{A}\left ( fF_z \right ) |_{z=z_1}dxdy}_{{\mathcal{S}_1}} - \underbrace{\int_{A}\left ( fF_z \right ) |_{z=z_0}dxdy}_{{\mathcal{S}_0}}  = \underbrace{\int_{V}\left ( E \frac{\partial f}{\partial t} + \mathbf{F} \cdot \nabla f \right ) dxdydz }_{{\mathcal{S}_R}}
+\underbrace{\int_{V} \left [ -f(W_{lr}+Q_{res}) \right ] dxdydz}_{{\mathcal{S}_D}}
-\displaystyle\int_{V} \frac{\partial\left ( fE \right )}{\partial t} dxdydz
\label{eq:cross_energy}
\end{equation}
where $E=B^2/8\pi$, $F_z=[v_zB_h^2-B_z(v_xB_x+v_yB_y)]/4\pi$, $W_{lr}$, $Q_{res}$ are the magnetic energy, vertical component of Poynting flux, work done by Lorentz force, and resistive heating, respectively.
The terms ${\mathcal{S}_0}$ and ${\mathcal{S}_1}$ represent net magnetic energy fluxes in the open regions at $z=z_0$ and $z=z_1$, respectively.
The term ${\mathcal{S}_R}$ is the magnetic energy flux transferred from the closed to open regions, and the term ${\mathcal{S}_D}$ is the energy dissipated through Lorentz force work and resistive heating.
The last term describes temporal changes in magnetic energy density in open regions and is negligible after time-averaging.
As in the mass flux calculations, we set $z_0 = 5$ Mm and $z_1 = 27$ Mm.
%The temporal evolution of ${\mathcal{S}_0}$, ${\mathcal{S}_1}$, ${\mathcal{S}_R}$, and ${\mathcal{S}_D}$ is presented in \fig{fig:mass_energy_flux_cross}(b), and the temporal average values of these terms is summarized in \fig{fig:cartoon}(b).
%\todo{This method will be employed to investigate the balance of the energy flux in Sect XXX.}

%\begin{equation}
%\nabla \cdot \left ( f \mathbf{F} \right ) = \left ( E \frac{\partial}{\partial t}+ \mathbf{F} \cdot \nabla \right )f -f(W_{lr}+Q_{res}) -\frac{\partial\left ( fE \right )}{\partial t}\nonumber
%\end{equation}

%\subsection{The role of chromospheric activities}

%Surges in our simulations move up and down, most of them fall back and failed to reach above 25 Mm.

%The strongest upward mass flux at high heights are contributed from the jets triggered in the corona \fig{fig:mass_flux_avg}.
%The density of coronal jets is lower, allowing them to reach higher heights.

%spicules are not resolved in our simulations.

%\todo{[I think there should be a better description of what is shown in Fig. 6 and in particular some discussion on the large variability and why we think that we can still trust the small averages despite all the variability. Also, there should be a clear definition of what is plotted here. All this is based on the Iijima paper, then Eqs (1) and (2), right? Then you should write explicitly which terms in those Eqs are plotted here.]}

Using the method outlined above, we calculate the mass and energy fluxed along open fields and the flux transfer from closed to open fields through interchange reconnection.
{We computed the values for datasets with both low and high temporal cadence of the writeout.
The low-cadence dataset spans several hours, providing an overview of mass transport trends, while the high-cadence dataset covers $\sim$20 minutes, offering more precise calculation of temporal derivatives but being more sensitive to individual jet events with lifetimes of a few hundred seconds.}
%The temporal evolution of ${\mathcal{M}_0}$, ${\mathcal{M}_1}$, and ${\mathcal{M}_R}$ is shown in \fig{fig:mass_energy_flux_cross}(a), and the temporal average values of these terms is summarized in \fig{fig:cartoon}(a).
The temporal evolution of the different terms in \eqn{eq:cross_mass} and \eqn{eq:cross_energy} is shown in \fig{fig:mass_energy_flux_cross} {for the low-cadence ($\sim$5.5 minutes) dataset and in \fig{fig:mass_energy_flux_cross_high_cad} for the high-cadence (10 seconds) dataset}.
{By comparing the two figures, we find that both datasets yield similar results in overlapping periods.
%, confirming the robustness of our approach.
}
The contribution of interchange reconnection to both mass flux ($\mathcal{M}_R$) and magnetic energy flux ($\mathcal{S}_R$) are mostly positive, indicating a continuous transfer of mass and energy from closed to open magnetic fields.
Moreover, these contributions are comparable to or even larger than the fluxes measured at $z = 27$ Mm ($\mathcal{M}_1$ and $\mathcal{S}_1$) at almost all times.
{Thus, we focus on the low-cadence dataset in this study to obtain the overall behavior of our simulations.}

It is worth mentioning that the mass flux term $\mathcal{M}_0$ exhibits significant temporal variability due to localized, dense chromospheric plasma intermittently moving up and down across $z = 5$ Mm within the open regions.
However, since the uncertainty in mass flux at different heights for a given snapshot is {largely} determined by the accuracy of field line tracing,
our use of a high-precision RK45 algorithm and oversampling by a factor of nine in the horizontal directions has {substantially reduced} these uncertainties. 
As a result, the calculated net mass flux in the open regions at any given time {appears reasonably} robust, despite the variability at lower heights.

Then we calculate the time-averaged fluxes of the different terms, and they are summarized in \fig{fig:cartoon}.
The uncertainties are estimated as the standard deviations of different snapshots divided by the square root of the number of snapshots.
The net imbalance primarily comes from the limited cadence of the dataset.
As already shown in \fig{fig:mass_flux},
the average mass flux in the open regions is negative at $z=5$ Mm and positive at $z=27$ Mm.
On average, $\langle{\mathcal{M}_R}\rangle$ is approximately 2.7 times larger than $\langle{\mathcal{M}_1}\rangle$, where bracket $\langle\cdot\rangle$ represents average over time.
It indicates that the mass supplied by interchange reconnection significantly exceeds the mass flux at greater heights.
Although it is challenging to determine how much of the chromospheric plasma ultimately escapes into the solar wind without passive tracers, the time-averaged net mass flux at $z=5$ Mm remains negative.
{For further comparison, the temporal evolution of the total mass flux at $z=5$~Mm is shown in \fig{fig:flux_whole_domain_5Mm}.}
{The time-averaged total mass flux at $z=5$ Mm is $(0.8\pm2.6)\times10^{-10}$~g~cm$^{-2}$~s$^{-1}$.
While its discrepancy with $\langle{\mathcal{M}_{1}}\rangle$ can be attributed to the limited temporal resolution of the dataset, the difference from $\langle{\mathcal{M}_{0}}\rangle$ still suggests the potential role of interchange reconnection.
}

%The average Poynting flux in the open regions is $(4.3\pm0.1)\times10^{5}$~erg~s$^{-1}$~cm$^{-2}$ at $z=5$ Mm and $(0.9\pm0.03)\times10^{5}$~erg~s$^{-1}$~cm$^{-2}$ at $z=27$ Mm.
%The contribution from interchange reconnection in between is $(2.1\pm0.2)\times10^{5}$~erg~s$^{-1}$~cm$^{-2}$, and the dissipated magnetic energy flux is $(5.0\pm0.1)\times10^{5}$~erg~s$^{-1}$~cm$^{-2}$.%$0.5\times10^{5}$~erg~s$^{-1}$~cm$^{-2}$, 
%The magnetic energy transport in the open field regions is also summarized in \fig{fig:cartoon}.

%\todo{[HP: not sure about this section. What is the purpose? Just to say that 1/3 is from interchange reconnection? But then you would have to provide/discuss evidence for this...]}\\
For magnetic energy flux, at $z = 5$ Mm, ${\mathcal{S}_0}$ is always positive, indicating continuous energy injection from below.
The ratio between $\langle{\mathcal{S}_0}\rangle$ and $\langle{\mathcal{S}_R}\rangle$ is approximately two, suggesting that about one-third of the Poynting flux in open magnetic fields above $z=5$ Mm originates from interchange reconnection,
while the remaining two-third is injected from below, likely in the form of Alfv\'enic waves.
These waves are thought to be driven by photospheric motions \citep{2022A&A...665A.118F} and/or by twist transport from closed loops to open field lines during magnetic reconnection \citep{2014SoPh..289.3043L}.
This result is consistent with \citet{2023ApJ...951L..47I}, where approximately half of the Poynting flux was attributed to interchange reconnection.
Still, $\langle{\mathcal{S}_R}\rangle$ is around 2.3 times higher than $\langle{\mathcal{S}_1}\rangle$, implying that interchange reconnection plays a role in supplying magnetic energy flux to the solar wind.
{Besides, we present the temporal evolution of the total magnetic energy flux at $z=5$~Mm, with a time-averaged value of $(11.7\pm0.4)\times10^{5}$~erg~cm$^{-2}$~s$^{-1}$.
This corresponds to nearly 30\% of the magnetic energy flux at this height in the closed loops being transported into open regions, comparable to the value ($\sim$20\%) in \citet{2023ApJ...951L..47I}.}
{It is worth mentioning that including more physics in the chromosphere, such as non-equilibrium ionization or ion-neutral interactions, may change the phenomena in the chromosphere and the magnetic free energy in the corona (through magnetic energy loss or flux associated with these processes), and possibly affect the values of the terms in \fig{fig:cartoon}. However, it would not remove the contribution from the interchange reconnection at coronal heights.}

Thus, while interchange reconnection is the dominant mechanism for mass loading onto open field lines at coronal heights,
magnetic energy in the corona is supplied through both wave propagation from the lower atmosphere and interchange reconnection within the corona, with their contributions being comparable.
Most of the magnetic energy dissipates within our calculation domain through Lorentz force work and Joule heating. 
{The dissipated energy may be partitioned into radiative losses, thermal conduction, kinetic energy, gravity potential of outgoing mass flow, among others, thereby} sustaining a 1 MK corona and driving a continuous outflow.

%>>>>>>>>>>>>>>>>>>>>>>>>>>>>>>>>>>>>>>>>>>>>>>>>>>>>>>>>>>>>>>>>>>>>>>>>>>>>>>
%\begin{figure*}[ht]
%\centering {\includegraphics[width=14cm]{mass_flux_avg_hor.pdf}} 
%\caption{Mass flux in open magnetic fields averaged over time and horizontal directions.
%} 
%\label{fig:mass_flux_avg}
%\end{figure*}
%<<<<<<<<<<<<<<<<<<<<<<<<<<<<<<<<<<<<<<<<<<<<<<<<<<<<<<<<<<<<<<<<<<<<<<<<<<<<<<

\end{appendix}

\end{document}